\documentclass[prb,aps,twocolumn,groupedaddress,floats,showpacs]{revtex4}
\usepackage{graphicx}
\usepackage{dcolumn}
\usepackage{bm}
\def\he4{$^4$He}
\def\Am3{\AA$^{-3}$}
\def\Am2{\AA$^{-2}$}
\def\beq{\begin{equation}}
\def\eeq{\end{equation}}
\begin{document}

\title{On the existence of  supersolid \he4 monolayer films}
\author{Massimo Boninsegni}
\affiliation{Department of Physics, University of Alberta,
Edmonton, Alberta T6G 2J1}
\date{\today}
\begin{abstract}
Extensive Monte Carlo simulations of \he4 monolayer films adsorbed on weak substrates have been carried out, aimed at ascertaining the possible occurrence of a quasi-two-dimensional supersolid phase. Only crystalline films {\it not} registered with underlying substrates are considered. Numerical results yield strong evidence that \he4 will {\it not} form a supersolid film on {any} substrate strong enough to stabilize a crystalline layer. On weaker substrates, continuous growth of a liquid film takes place. 
\end{abstract}

 \maketitle
\section{Introduction}
The search for a supersolid phase of \he4 is still ongoing, not only in the
bulk crystal, but also in films of helium adsorbed on different substrates. 
Indeed, long before the observation of non-classical rotational inertia in 
solid \he4 by Kim and Chan,\cite{KC}  it was proposed by Crowell and 
Reppy \cite{crowell} that the second layer of 
a \he4 film adsorbed on graphite might display simultaneously superfluid
and solid behaviour; this contention has been recently re-iterated.\cite{fuku}
The most recent, accurate theoretical studies cast serious doubts on 
this claim;\cite{corboz,dang} furthermore, one should note that the denomination ``supersolid" 
is scarcely appropriate for such a system, as the (alleged) crystalline order of the second 
layer would not arise spontaneously as required by the definition, but rather be 
induced by an external agent, namely the graphite substrate. More generally, any 
claim of supersolid behaviour made in the context of a quantum film, requires 
that the crystalline phase of the latter be {\it incommensurate} with the substrate upon which 
the film is adsorbed. 
\\ \indent
The idea that, in a  many-body system wherein particle motion is almost entirely confined to two dimensions,  quantum fluctuations in the transverse direction may give rise to a superfluid response, even  if  the effective two-dimensional (2D) particle density corresponds to a crystalline phase in the strictly 2D system, has already been extensively explored in the context of the search for a superfluid phase of parahydrogen. However, first principles calculations carried out for various substrates \cite{melting,joe} did not yield any evidence of superfluidity, a fact that could be certainly ascribed to the strong tendency of parahydrogen to solidify, which renders it qualitatively different from helium. It is therefore not implausible that the more quantal helium may display a different behaviour in similar physical conditions.
\\ \indent
The possible supersolid behavior of quasi-two-dimensional helium, with
motion of atoms in one direction  restricted by a confining potential, was investigated 
in a recent theoretical study, in which a simple harmonic model of transverse confinement 
was adopted.\cite{cazzorla} The results of that study were interpreted by the authors as suggestive of a 
novel phase, simultaneously displaying a superfluid response and broken 
translational invariance, underlain by large atomic excursions in the transverse direction.  
\\ \indent
As the harmonic approximation may reasonably well describe the
potential experienced by helium atoms in the topmost layer of a thin film, or in the first adsorbed layer on a 
weak substrate, the question naturally arises  of whether
there exists an actual experimental system that could give rise to observable quasi-two-dimensional supersolid \he4. 
Viable candidates are  the second incommensurate solid \he4 layer on graphite,\cite{pierce}
or perhaps the first layer on  a weak 
substrate,  barely strong enough  to stabilize it; the latter scenario is explored in detail here.

This paper illustrates results of  Quantum Monte Carlo (QMC) simulations of an ensemble of $N$ $^4$He atoms, regarded as point particles,  moving in the presence of an infinite, smooth planar substrate of area $A$, positioned at $z$=0. The nominal $^4$He coverage is $\theta=N/A$. 
The quantum-mechanical many-body  Hamiltonian  is the following:
\begin{equation}\label{one}
\hat H = -{\lambda}\sum_{i=1}^N \nabla_i^2 + \sum_{i<j} V(r_{ij}) +\sum_{i=1}^N U(z_i)\ .
\end{equation}
Here, $\lambda$=6.0596 K\AA$^2$, $V$ is the potential describing the interaction between two helium atoms,  only depending on their relative distance, whereas $U$ is the potential describing the interaction of a helium atom with the substrate,  also depending only on the distance of the atom from the substrate. We use the accepted Aziz potential\cite{aziz79} to describe the interaction of two helium atoms. \\ \indent
The interaction of a helium atom with the substrate (i.e., the $U$ term in (\ref{one})) is described here by means of the
so-called ``3-9" potential:
\begin{equation}\label{39}
U_{3-9}(z) = \frac{D}{2}\biggl ( \frac {a^9}{z^9}-3\frac{a^3}{z^3}\biggr )
\end{equation}
which is a functional form obtained by integrating the Lennard-Jones potential over a semi-infinite, continuous slab. Here, $D$ is the characteristic depth of the attractive well of the potential, whereas $a$ is essentially the distance from the plane of the minimum of such a well; the potential is strongly repulsive for $z < a$, weakly attractive for $z > a$. 

The aim of $U_{3-9}$ is that of providing a fairly realistic description of the environment experienced by a helium atom in the vicinity of a relatively weak substrate.  Substrate corrugation, which is  physically responsible for the appearance of low-coverage insulating films {\it registered} with the crystalline structure of the  underlying substrate, is purposefully omitted in our model, as we are only interested in {\it incommensurate} solid films. 
\\ \indent
For comparison purposes, we also performed calculations using the same harmonic potential of Ref. \onlinecite{cazzorla}, namely
\begin{equation}\label{ho}
U_{H}=\frac{\lambda\ (z-z_\circ)^2}{4\ \sigma^4}
\end{equation}  
defined in terms of the root-mean-square excursion $\sigma$ away from the equilibrium position (located at $z_\circ$ above the plane) of the physisorption potential experienced by a free $^4$He atom near a given substrate. Obviously, $\sigma$ is the only parameter in this model, the actual value of $z_\circ$ being immaterial.

The low-temperature thermodynamics of the many-body system described by Eq. (\ref{one}) has been investigated by QMC simulations based on  the continuous-space worm
algorithm, a methodology that has proven quite effective in the study of
Bose systems. Details of the computational methodology adopted in this work are described in the literature.\cite{worm,worm2,cuervo} The results presented here are based on simulations of systems comprising $N$=144 atoms; the imaginary time step utilized in most of the calculations is $\tau$ = 1.5625$\times 10^{-3}$  K$^{-1}$, which has been empirically found to provide numerical estimates indistinguishable, within statistical errors, from those extrapolated to the $\tau=0$ limit. The substrate geometry is chosen rectangular in all simulations, so as to accommodate a triangular crystal. Periodic boundary conditions are assumed in all directions, but the size of the system in the $z$ direction (i.e., perpendicular to the substrate) is chosen large enough to make boundary conditions unimportant.  This set-up is standard for QMC simulations of adsorbed films.\cite{glass}

Our main result is that no quasi-two-dimensional supersolid \he4 phase arises in the physical conditions described above. Specifically, strong atomic  transverse confinement (corresponding either to a deep attractive well $D$ in $U_{3-9}$ or to a low value of $\sigma$ in $U_H$)  leads to the formation of non-superfluid crystalline monolayer films, whereas on weaker substrates continuous growth of superfluid (liquid)  film is observed. No intermediate homogeneous phase, simultaneously displaying superfluidity and broken translational invariance, is ever observed in our simulations, including any metastable ``glassy" superfluid phase.  These results and conclusion are largely consistent with those of Ref. \onlinecite{cazzorla}; in our view, however,  any interpretation of the available numerical evidence  in terms of possible supersolidity, is baseless.

In the remainder of this manuscript we illustrate our results, discussing first those  for for the more realistic $U_{3-9}$, and then the ones yielded by the simple confining potential $U_H$; as we shall see, there exist no substantial qualitative differences between the physical behaviours observed in two cases.
\section{``3-9" potential}

Table \ref{tab:1} 
shows values of the coefficients $a$ and $D$ which have been proposed (and utilized in previous works) to describe the interaction of a helium atom with various metal substrates,\cite{vidali,chiz} based on the  form (\ref{39}) of the physisorption potential. The most attractive of them is Al, for which the well depth $D$ is 60.3 K, the least attractive is Li  ($D=17.9$ K). It should be stated at the outset that, aside from the very simple form of the ``3-9" model potential, these coefficients are only known to a typical accuracy of 10-20\%. The purpose of using a simple model such as (\ref{39}), even with the current uncertainty in the values of the coefficients, is that of gaining fundamental understanding on the physics of adsorbed monolayer and guiding the experimental search toward systems that may display novel behaviour.

\begin{table}
\caption{Values of the coefficients $a$ and $D$ of the ``3-9" potential describing the interaction between a helium atom and four metal substrates, listed in order of decreasing attractiveness (i.e., the well depth $D$).}
\label{tab:1}       
\begin{tabular}{c|c|c|c}
\hline\noalign{\smallskip}
Substrate &Reference & $a$ (\AA) & $D$ (K)  \\
\noalign{\smallskip}\hline\noalign{\smallskip}
Al  & \onlinecite{vidali} &2.96 & 60.3 \\
Ni &  \onlinecite{vidali} & 3.26 & 48.7 \\
Mg &  \onlinecite{chiz} &3.26  &35.6 \\
Li & \onlinecite{chiz} &3.76 &17.9\\
\noalign{\smallskip}\hline
\end{tabular}
\end{table}

Calculations making use of the ``3-9" potential were all carried out at the same coverage $\theta=0.08$ \Am2, chosen  high enough to fall  unambiguously into the crystal region of the two-dimensional (2D) phase diagram of \he4 at low $T$,\cite{gordillo} but also sufficiently low to make transverse zero-point motion significant. We have carried out calculations at low $T$, the lowest being $T=0.1$ K. In general, results obtained at $T\lesssim 0.5$ K can be regarded as essentially ground state estimates, as no significant temperature dependence can be observed at lower $T$. 

\begin{figure}[tbp]
\centerline{\includegraphics [angle=-90, scale=0.33]{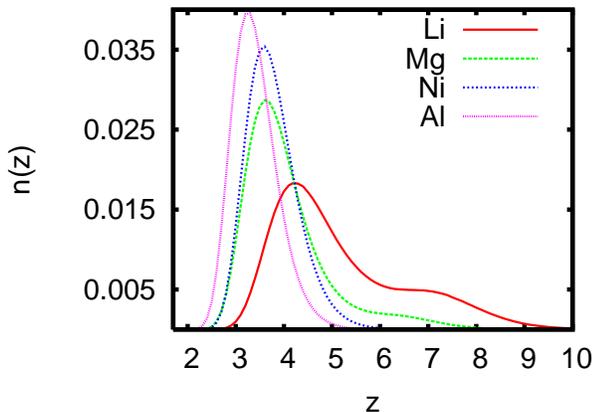}} 
\caption{(Color online).
\he4 density profile $n(z)$ (in \AA$^{-3}$) of \he4 in the direction perpendicular to the substrate, computed at $T$=0.2 K for the four substrates listed in Table \ref{tab:1}. The nominal coverage is $\theta$=0.08 \Am2 in all cases.} \label{f1}
\end{figure}

Figure \ref{f1} shows computed \he4 density profiles $n(z)$ in the direction ($z$) perpendicular to the substrate, for the four metal substrates listed in Table \ref{tab:1}. Specifically,
\begin{equation}
n(z)\equiv ({1/A}) \int dx\ dy \ \rho(x,y,z)\ ,
\end{equation} 
where $\rho$ is the three-dimensional \he4 density.  

The most weakly attractive of the four substrates, namely Li, is the weakest known substrate on which \he4 is predicted to form a stable superfluid monolayer, at a coverage of approximately 0.056 \Am2. On increasing coverage, promotion of \he4 atoms to second layer is observed before monolayer crystallization,\cite{boninsegni1,boninsegni2} as seen in Fig. \ref{f1} from the broadly extended shape of $n(z)$, which features a shoulder beyond its main peak at $z\approx$ 4.5 \AA. The same physics is observed on a Mg substrate, only slightly more attractive than Li. On Ni and Al, on the other hand, $n(z)$ displays a single peak, suggesting that \he4 atoms form essentially a {\sl monolayer}, positioned closer to the substrate for the more attractive Al, with significant zero-point motion of atoms in the $z$ direction.
\begin{figure}[tbp]
\centerline{\includegraphics [angle=-90, scale=0.33]{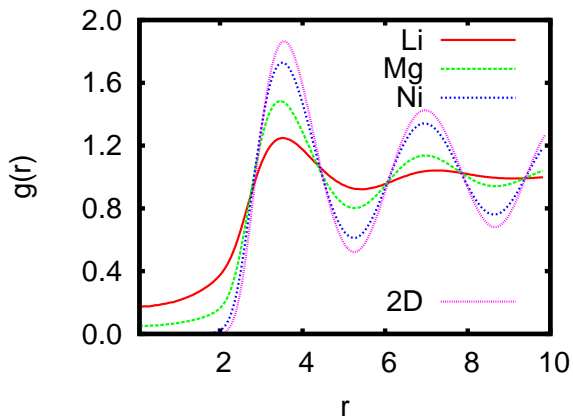}} 
\caption{(Color online).
Reduced \he4 pair correlation function, defined as in Eq. (\ref{gr}), at $T$=0.2 K for three of the four substrates listed in Table \ref{tab:1} (result for Al is indistinguishable from that for Ni, on the scale of the figure). The nominal coverage is $\theta$=0.08 \Am2 in all cases. Also shown (dotted line) is the pair correlation function for two-dimensional \he4 at the same temperature and coverage.} \label{f2}
\end{figure}

The 2D character of \he4 films on Ni and Al can be established by looking at the the angularly averaged, �reduced� pair correlation function $g(r)$, with $r = \sqrt{x^2+y^2}$ and
\beq\label{gr}
g(x,y) = \frac{1}{A\theta^2}\ \int dx^\prime\ dy^\prime\ n(x+x^\prime, y+y^\prime)\ n(x^\prime,y^\prime)
\eeq
with $n(x,y)\equiv \int dz\ \rho(x,y,z)$. The more closely $g(r)$ mimics the pair correlation function of a strictly 2D \he4 system of the same coverage, the closer the film is to being essentially 2D. Figure \ref{f2} shows $g(r)$ for a \he4 film of coverage $\theta$, adsorbed on Li, Mg and Ni substrates, compared to the pair correlation function of \he4 in two dimensions. 
As mentioned above, in two dimensions \he4 at this coverage forms an insulating (i.e., non superfluid)  triangular crystal at low $T$, and that is the physics observed on the two stronger substrates, namely Al and Ni, for which $g(r)$ closely approaches the 2D shape. Atomic zero-point motion in the direction perpendicular to the substrate only causes a small depression of the main peak. On the other hand, on the weakly attractive Li and Mg, promotion of \he4 atoms to the second layer is shown by the finite value that $g(r)$ takes at the origin, as well as by the less pronounced first peak and oscillations. On these two substrates, no 2D crystal of helium forms, and the \he4 superfluid fraction is essentially 100\% at $T\lesssim 0.5$ K, whereas no detectable superfluid signal is observed on Ni and Al, down to a temperature $T$=0.1 K. 
\begin{figure}[tbp]
\centerline{\includegraphics [angle=-90, scale=0.33]{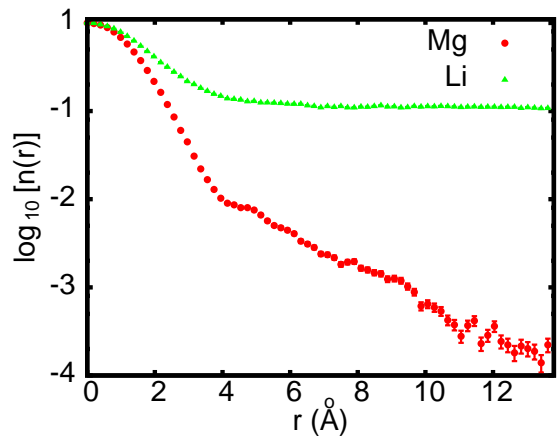}} 
\caption{(Color online).
One-body density matrix $n(r)$, computed at $T$=0.2 K for Li (triangles) and Ni (circles) substrates. The nominal coverage is $\theta$=0.08 \Am2, whereas the temperature is $T$=0.2 K. No significant change is  seen at lower $T$.} \label{f3}
\end{figure}
 These two very different behaviours are also reflected in the one-body density matrix $n(r)$ (shown in Fig. \ref{f3}), which displays an exponential decay for the film adsorbed on Ni and Al, whereas on Mg and Li its expected much slower power-law decay is barely detectable within the precision of our calculation.\cite{npteo} 

Looking at Table \ref{tab:1}, we see that the value of $a$ for Mg and Ni substrates is the same, the different physics of adsorbed helium films arising exclusively from the different well depths (i.e., $D$). The question can be posed, therefore, of whether on a substrate characterized by that very same value of $a$ and one of $D$ between those of Mg and Li, a ``supersolid" \he4 monolayer may exist, namely an effectively 2D crystalline film, featuring a finite superfluid response, possibly within a narrow range of temperature. If one could establish that supersolid behaviour can be observed at an attainable low $T$ a narrow range of values of $D$, then issue would of course be that of fashioning such a substrate in the laboratory. One possibility might be that of ``weakening" an attractive substrate by pre-plating it with a rare gas, such as Kr.\cite{wie}
\begin{figure}[tbp]
\centerline{\includegraphics [angle=-90, scale=0.33]{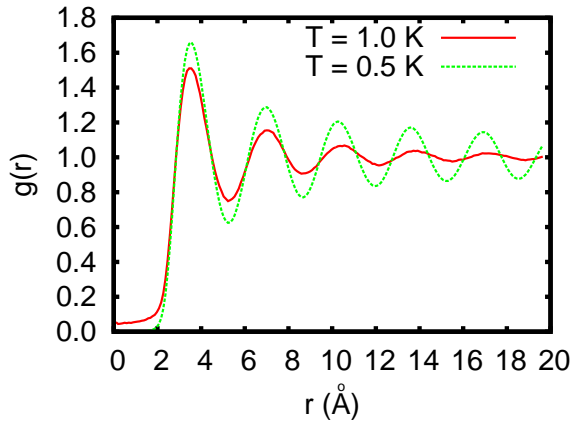}} 
\caption{(Color online).
Integrated pair correlation function computed at $T$=1.0 (solid line) and 0.5 (dashed line) K, for a \he4 film of coverage $\theta$=0.080 \Am2 on a ``3-9" substrate with $D$=40 K and $a$=3.26 \AA.} \label{f4}
\end{figure}

We have addressed this issue by performing the same calculations for a ``fictitious" substrate of well depth $D$ in the range between 36 and 40 K and with $a$=3.26 \AA, in search of a ``sweet spot" wherein quantum oscillations in the transverse direction may give rise to novel physics. As shown below, however, no other phase is observed other than a (non-superfluid) solid and a (superfluid) liquid film. \\ \indent
Figure \ref{f4} shows the integrated pair correlation function $g(r)$ computed on such a model substrate, with $D$=40 K, at two low temperatures, namely $T$=1 K and $T$=0.5 K. It is worth mentioning that the result at $T$=0.75 K is indistinguishable, within the statistical errors of the calculation, from that at $T$=0.5 K. Clearly, the physical behaviour of the system changes qualitatively as the temperature is lowered. At the lower $T$, the pair correlation function features the regular oscillations proper of a crystal, and is essentially 2D in character; at the higher temperature, on the other hand, little or no structure appears beyond the first peak, and the result is consistent with liquid-like behaviour.  Moreover, in the latter case the small but finite value of $g$ at the origin points to thermally activated promotion of helium atoms to second layer, i.e., the quasi-2D crystal melts into a superfluid at  a temperature $T_m \approx 1$ K. 
\begin{figure}[tbp]
\centerline{\includegraphics [angle=-90, scale=0.33]{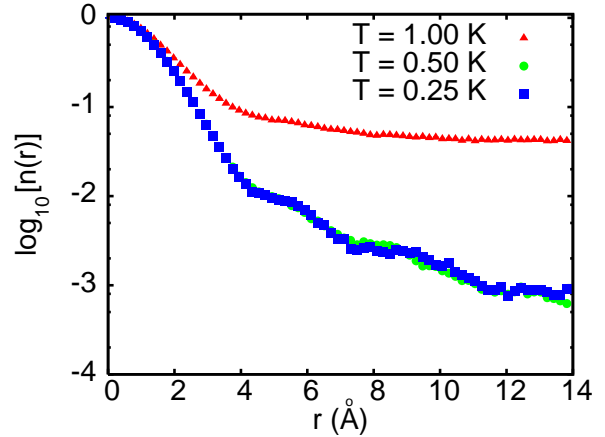}} 
\caption{(Color online).
One-body density matrix $n(r)$, computed at $T$=1.0 K (triangles), $T$=0.5 K (circles) and $T$=0.25 K (boxes), for a \he4 film of coverage $\theta$=0.080 \Am2 on a ``3-9" substrate with $D$=40 K and $a$=3.26 \AA. Statistical errors are of the order of the sizes of the symbols.}\label{f5}
\end{figure}

The superfluid character of the film at higher temperature can be established by direct computation of the superfluid density, by means of the well-known ``winding number" estimator, as well as through the behaviour of the integrated one-body density matrix, shown in Figure \ref{f5}. While at the higher $T$ the function displays a slow decay (and correspondingly the value of the superfluid density is approximately 50\% for the finite system simulated), at lower $T$ it decays exponentially, with no appreciable dependence on $T$, as shown in Figure \ref{f5}. In fact, the result at $T$=0.25 K is indistinguishable from that at $T$=0.5 K, within statistical errors. \\ \indent
The above results represent the general trend observed on performing calculations with {\it any} value of $D$, for fixed $a$. Increasing (decreasing) $D$ has simply the effect of raising (lowering) the temperature $T_m$ at which the quasi-2D crystal melts. If $D$ is large, the crystal melts into a normal liquid, whereas superfluidity is observed for intermediate values of $D$. On the other hand, as $D$ approaches a value close to $\sim$ 36 K, i.e. near that corresponding to a Mg substrate, then $T_m\to 0$, i.e., no crystal film is observed. Simultaneous occurrence of superfluid and crystalline properties is {\it never} observed.\\ \indent
Thus, the physical conclusion of this theoretical exercise, is that quantum-mechanical particle fluctuations in the transverse direction do {\it not} lead to the stabilization of a quasi-2D supersolid phase of \he4, at least within the framework defined by a potential such as (\ref{39}).
\section{Harmonic potential}
The conclusion at which we arrived in the previous section would appear to clash with recent predictions of supersolid behavior from Cazorla {\it et al.} (Ref. \onlinecite{cazzorla}), who performed a strictly $T$=0 calculation based on a harmonic model of confinement (Eq. \ref{ho}). It seems difficult to ascribe such a different physical outcome to the  functional form the harmonic potential, which should provide an excellent approximation to any realistic physisorption potential, especially one strong enough to allow only relatively small quantum excursions in the transverse direction. 
\\ \indent
In order to obtain an independent verification of the predictions of Ref. \onlinecite{cazzorla}, we have carried out  finite temperature calculations for the harmonic model (\ref{ho}) of confining potential, using the same values of the parameter $\sigma$ in Eq. (\ref{ho}), as well as of coverage $\theta$, as in Ref. \onlinecite{cazzorla}. 
\begin{figure}[tbp]
\centerline{\includegraphics [angle=-90, scale=0.33]{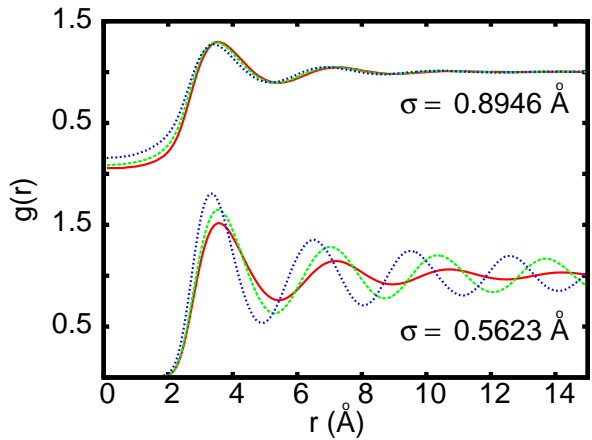}} 
\caption{(Color online).
Integrated pair correlation function computed at $T$=1.0 for a \he4 film of coverage $\theta$=0.0719 \Am2 (solid line), 0.0788 \Am2 (dashed line) and 0.0918 \Am2 (dotted line), using the harmonic approximation (Eq. \ref{ho}) for the substrate-adatom interaction. The two sets of data refer to two values of the parameter $\sigma$, yielding the root-mean-square excursions of free atoms in the transverse direction. } \label{f6}
\end{figure}
Figure \ref{f6} displays integrated pair correlation functions for two different values of the parameter $\sigma$ of Eq. \ref {ho} and the three different coverages. Results shown pertain to simulations at a temperature $T$=1.0 K, but calculations performed at lower temperature (down to 0.1 K) yield indistinguishable results, within statistical uncertainties. 

Two different regimes can be straightforwardly identified, which correspond to the same already observed using the ``3-9" potential. 
Specifically, for the greater value of $\sigma$, corresponding to looser transverse confinement, the helium films do not display a 2D character. Indeed, as shown by the curves in the upper part of Figure \ref{f6},  the pair correlation functions all remain finite in the region of the hard core of the interatomic potential, signalling particle layering. The superfluid fraction in all three cases is essentially  100\% at $T \lesssim 0.5$ K, i.e., the system features no broken translational invariance.\cite{leggett} No hint of crystalline structure can be seen. We conclude that the system in this case forms a multi-layered liquid, as condensed helium invariably does on weak substrates. In other words, the physical behavior observed here closely reproduces that seen on weak substrates, such as Mg. All of this is in agreement with the findings of  Ref. \onlinecite{cazzorla} for this value of $\sigma$, for which authors find that the ground state is a structureless liquid. 

The physics is clearly very different for the case of tighter confinement, represented by the lower value of $\sigma$, for which the behavior of the system is essentially 2D, atomic motion in the transverse direction being strongly suppressed. Indeed, the integrated pair correlation function closely resembles that of the purely 2D system, as we have verified by simulation.  For the two higher coverages, the $g(r)$ displays regular, robust oscillations characteristic of the crystalline phase; concomitantly, the value of the superfluid density is {\it zero}, as permutations of indistinguishable \he4 atoms are exceedingly infrequent, down to the lowest temperature explored here.
Ref. \onlinecite{cazzorla} reports minuscule ($\sim 10^{-4}$) but {\it finite} values of the superfluid fraction at $T$=0, for the same values of $\sigma$ and $\theta$ considered here. If those numbers could be regarded as  reliable bulk estimates, they would point to a supersolid phase for the two larger coverages, with a tiny superfluid response.\cite{boh} However, it is far from clear whether the computational methodology that yielded these estimates (Diffusion Monte Carlo)  truly affords that kind of precision. The robustness of these quoted values of the superfluid fraction  against various sources of systematic error  (finite size of the system, population size and  trial wave function bias, time step error), whose magnitude could easily dwarf  that of the reported superfluid signal, is not discussed at all in Ref. \onlinecite{cazzorla}.  
\\ \indent
For the lowest coverage, namely 0.0719 \Am2, the system is a quasi-2D superfluid, with no diagonal long-range order; the superfluid density in this case is finite, and saturates to unity in the low temperature limit. There is no evidence that the system may crystallize at sufficiently low $T$, for unlike the case of weak substrates explored above, the liquid character here is not associated to the formation of multiple layers. 
Our numerical prediction for the lowest coverage case, therefore, is in  disagreement with that of Ref. \onlinecite{cazzorla}, wherein a crystalline ground state at all coverages is predicted,\cite{kemerda} for $\sigma$=0.5623 \AA.
\\  \indent A few remarks are in order.
First, the most recent study\cite{gordillo} of the low temperature phase diagram of 2D \he4 (based on a slightly more attractive version of the Aziz potential and smaller system sizes than those utilized here), places this value of coverage inside the very narrow coexistence region of liquid and solid phases;  this  clearly renders problematic the unambiguous identification of a single phase, especially one displaying two kinds of order. One would expect, however, that in the presence of competing liquid and crystalline phases of very similar energy in 2D,  transverse fluctuations should strengthen the disordered phase. 
Moreover,  the prediction of Ref. \onlinecite{cazzorla} is based exclusively on the resolution of an expectedly tiny energy difference (values are not given in Ref. \onlinecite{cazzorla}). It is worth reminding that finite temperature methods are unbiased, i.e., they require no {\it a priori} physical assumption. On the other hand, the values of physical quantities computed with ground state projection methods such as the one used in Ref. \onlinecite{cazzorla}, {\it including the energy}, are inherently biased by the use of initial trial wave functions.  
\section{Conclusions}
We have carried out extensive numerical studies of \he4 films adsorbed on weak substrates, searching for a possible quasi-2D supersolid phase, i.e., one displaying simultaneously diagonal and off-diagonal long range order. We have restricted our search to the case in which density long-range order occurs through {\it spontaneous} breaking of translational invariance, i.e., it is not caused by an external pinning potential.  \\ \indent
The hypothesis tested here is that quantum fluctuations of atoms in the direction transverse to the plane may enhance the tendency of identical particles to exchange, possibly leading to the global superfluid phase coherence, while preserving density long-range order. The results shown here, obtained for two different models of substrate adsorption,  constitute strong evidence that such mechanism does not take place; only non-superfluid crystalline phases or superfluid liquid ones are observed, depending on the weakness of the substrate and on the temperature. It seems, therefore, that this physical approach, namely effective reduction of dimensionality, is not likely to lead to the observation of a supersolid phase.
\section*{Acknowledgments}
This work was supported by the Natural Science and Engineering Research
Council of Canada under grant G121210893. Hospitality of the Ecole Normale Sup\'erieure de Lyon (France), where most of the work was carried out, is gratefully acknowledged.


\begin{thebibliography}{99}

\bibitem{KC} E. Kim and M. H. W. Chan, Nature,
{\bf 427}, 225 (2004); Science {\bf 305}, 1941 (2004).

\bibitem{crowell}
 P. A. Crowell and J. D. Reppy, Phys. Rev. Lett. {\bf 70}, 3291 (1993).

\bibitem{fuku}
Y. Shibayama, H. Fukuyama, and K. Shirahama, J. Phys.: Conf.
Ser. {\bf 150}, 032096 (2009).
 
 \bibitem{corboz}
 P. Corboz, M. Boninsegni, L. Pollet and M. Troyer, 
Phys. Rev. B {\bf 78}, 245414 (2008).

\bibitem{dang}
L. Dang and M. Boninsegni, 
Phys. Rev. B {\bf 81}, 224502 (2010).

\bibitem{melting}
M. Boninsegni,
Phys. Rev. B {\bf 70}, 125405 (2004). 

\bibitem{joe}
J. Turnbull and M. Boninsegni,
Phys. Rev. B {\bf 76}, 104524 (2007). 

\bibitem{cazzorla}
C. Cazorla, G. E. Astrakharchik, J. Casulleras and J. Boronat,
J. Phys.: Condens. Matter {\bf 22}, 165402 (2010).

\bibitem{pierce}
Successive layers, on graphite as well as on other attractive substrates, such as glass, 
are believed to be liquid. See, for instance, refs. \onlinecite {manousakis, glass}.

\bibitem{manousakis} 
M. Pierce and E. Manousakis, Phys. Rev. B 
{\bf 63}, 144524 (2001). 

\bibitem{aziz79}
R. A. Aziz, V. P. S. Nain, S. Carley, W. L. Taylor and G. T.
McConville, J. Chem. Phys. {\bf 70}, 4330 (1979).


\bibitem{worm} M. Boninsegni, N. Prokof'ev, and B. Svistunov,
Phys. Rev. Lett. {\bf 96}, 070601 (2006).

\bibitem{worm2} M. Boninsegni, N. Prokof'ev, and B. Svistunov,
Phys. Rev. E {\bf 74}, 036701 (2006).

\bibitem{cuervo}
J. E. Cuervo, P.-N. Roy and M. Boninsegni, J. Chem. Phys. {\bf 122}, 114504 (2005).



\bibitem{glass}
M. Boninsegni, J. Low Temp. Phys. {\bf 159}, 441 (2010). 


\bibitem{vidali}
G. Vidali, G. Ihm, H.-Y. Kim, M.W. Cole, Surf. Sci. Rep. {\bf 12}, 133 (1991).

\bibitem{chiz}
A. Chizmeshya, M. W. Cole, and E. Zaremba, J. Low Temp.
Phys. {\bf 110}, 677 (1998).

\bibitem{gordillo}
M.-C. Gordillo and D. M. Ceperley, Phys. Rev. B {\bf 58}, 6447 (1998).


\bibitem{boninsegni1}
M. Boninsegni, M. W. Cole and F. Toigo, Phys. Rev. Lett. {\bf 83}, 2002 (1999).

\bibitem{boninsegni2}
M. Boninsegni and L. Szybisz, Phys. Rev. B {\bf 70}, 024512 (2004).

\bibitem{npteo}
The one-body density matrix computed here is defined as a three-dimensional object, which is cylindrically and translationally averaged. Because we are considering here films of one or two atomic layers, its long-range behaviour reflects 2D physics.

\bibitem{wie}
See, for instance, H. Wiechert, K. D. Kortmann, and N. St\"usser, Phys. Rev. B {\bf 70}, 125410 (2004).

\bibitem{leggett}
A. Leggett, Phys. Rev. Lett. {\bf 25}, 1543 (1970).


\bibitem{boh} It is puzzling that the authors of Ref. \onlinecite {cazzorla} refer to their proposed supersolid phases as ``glassy", given that their pair correlation functions are essentially the same with those of regular 2D crystals, displaying persistent oscillations for long inter-atomic separations. On the other hand, a glassy phase features short-range order only, much like a liquid, and indeed the pair correlation functions are essentially those of a liquid. See, for instance, M. Boninsegni, N. V. Prokof'ev and B. Svistunov, Phys. Rev. Lett. {\bf 96}, 105301 (2006).

\bibitem{kemerda}
The slight difference in the interatomic potential utilized in our study and theirs cannot account for the different physics observed. We carried out calculations using the same version of the Aziz potential used in Ref. \onlinecite{cazzorla}, and obtained the same physical results as with the potential of Ref. \onlinecite{aziz79}.
 \end{thebibliography}
\end{document}